\documentclass[twocolumn,showpacs]{revtex4}
\usepackage{epsf}
\usepackage{epsfig}
\usepackage{amsmath}
\usepackage{dcolumn}
\usepackage{graphicx}

\def\Vr{\mathbf r}
\def\VR{\mathbf R}
\def\VQ{\mathbf Q}
\def\Vk{\mathbf k}
\def\VF{\mathbf F}
\def\smhalf{{\textstyle \frac{1}{2}}}
\def\FT{\hat{\mathcal F}}

\def\be{\begin{eqnarray}}
\def\ee{\end{eqnarray}}
\def\beq{\begin{equation}}
\def\eeq{\end{equation}}
\newcommand{\tx}[1]{\mathrm{#1}}

\begin{document}
\draft
\title{Multiscale simulations in simple metals: a density-functional based
  methodology}
\author{Nicholas Choly$^{(1)}$\footnote{Electronic Address:
    {\tt choly@fas.harvard.edu}}, Gang Lu$^{(1)}$,
  Weinan E$^{(2)}$ and Efthimios Kaxiras$^{(1)}$}
\affiliation{$^{(1)}$Division of Engineering and Applied Sciences,
     Harvard University, Cambridge, MA 02138 \\
     $^{(2)}$Department of Mathematics and PACM, Princeton University,
     Princeton, NJ 08544}
\date{\today}

\begin{abstract}
We present a formalism for coupling a density functional theory-based
quantum simulation to a classical simulation for the treatment of simple
metallic systems.  The formalism is applicable to multiscale
simulations in which the part of the system requiring
quantum-mechanical treatment is spatially confined to a small region.
Such situations often arise in physical systems where chemical interactions
in a small region can affect the macroscopic mechanical  properties of
a metal.
We describe how this coupled treatment can be accomplished efficiently,
and we present a coupled simulation for a bulk aluminum system.
\end{abstract}

%

%\pacs{PACS Numbers: 77.80.Bh, 64.60.Cn, 77.84.Bw, 63.20.Dj}
%
%\vskip1pc]

\maketitle
%\twocolumn

\section{Introduction}
In performing computer simulations of complex physical systems, a
premium is placed on accuracy and efficiency.  Typically, one of
these qualities can be improved at the expense of the other.  In
recent years, a new approach has emerged that addresses a class of
problems in which important small-length-scale phenomena are
confined to a small region of the system but can have an impact on
the behavior over a much larger scale. A typical case is the tip
of a crack, where localized chemical reactions may affect the
strength of interatomic bonding, which in turn can influence in a
dramatic way the macroscopic mechanical properties of the solid.
Such problems fall under the rubric of ``multiscale'' phenomena,
requiring a treatment that addresses important aspects at each
scale.  The novel feature of this type of simulation is to use an
accurate but computationally demanding method to treat the region
of the system in which small-length-scale degrees of freedom are
important, and a faster but less accurate method with the
small-length-scale physics ``coarse-grained'', to treat the rest
of the system.

Multiscale approaches rely on successfully coupling the two (or
more) regions involved, which is referred to as {\em seamless
coupling}. There is no single notion as to what constitutes a
seamless coupling, but generally the coupling should be
accomplished in such a way that the fictitious boundary between
the two regions, which only exists in the coupled simulation and
not in the real system, does not introduce any physical
consequences.  For instance, recently several papers have dealt
with the issue of ensuring that phonons are not reflected by the
boundary between the two coupling methods\cite{WeinanH02,
CaidBY00}.  In the consummate multiscale method, the resulting
energetics or dynamics is indistinguishable from what would result
from a calculation with the accurate method applied to the {\em
entire} system.  This ideal would be achieved only if the two
simulation methods involved were seamlessly matched at the
boundary, and further, only if the part of the system treated by
the faster, less accurate method was indeed free of important
small-scale physics.  Another important but obvious characteristic
of a good multiscale method is that the computational overhead of
performing a coupled simulation is not significant. More
specifically, the computation time for the coupled simulation
should be on the order of computation time required for the
accurate method to treat the small detailed region, since the time
required for the less accurate method to treat the rest of the
system is typically several orders smaller; nothing is gained if
the coupling is so costly that the coupled method takes as long as
using the accurate method to treat the whole system.  When the
approach requires coupling a quantum mechanical method to a
classical method, additional complications arise because of the
presence of electronic degrees of freedom in the quantum
mechanical region; thus boundary conditions on the electron
wavefunctions must be imposed at the interface between the
regions.  Density functional theory (DFT) provides a significant
simplification over more direct quantum mechanical methods in that
the calculation of ground state energies and forces requires the
minimization of a functional of the electron density $\rho(\Vr)$
only\cite{HohenbergK64}. Thus, in principle boundary conditions
need only be imposed on $\rho(\Vr)$. This statement only applies
to the formulation of the problem that does not invoke the
explicit calculation of electronic wavefunctions (the most common
way of implementing DFT actually does involve individual
electronic wavefunctions, the so-called Kohn-Sham
orbitals\cite{KohnS65}). Coupling an approximate DFT calculation
that is based on the electronic density alone to a classical
interatomic potential should be more straightforward than coupling
an orbital-based quantum mechanical method to a classical method.

The present article describes a formalism for concurrently
coupling a system consisting of two regions, one treated with
density functional theory (without invoking electronic
wavefunctions) and the other with a classical interatomic
potential. Due to the type of approximations involved, the present
approach is particularly well suited for simple metallic systems;
we emphasize, however, that this is not an inherent limitation of
our approach, but rather a limitation imposed by the shortcomings
of the methodologies employed for the treatment of the various
parts of the system, and if these are eliminated the approach
could be generally applicable. In section \ref{sec:background},
other methodologies for coupling multiple simulation approaches
and their relevance to the present methods are discussed.  In
Section \ref{sec:formalism}, the formalism of the present class of
coupling methods is established. In Section
\ref{sec:implementation}, details of implementing the methods and
achieving efficiency are presented and some tests of the method
are reported in Section \ref{sec:tests}. Finally we conclude in
Section \ref{sec:conclusions} with a discussion of the results of
the tests.

\section{Background}\label{sec:background}

A large number of concurrent multiscale methods\cite{Gaoreview96,
GovindWdC98, BroughtonABK99} approach the problem of coupling two
different simulation methods by writing the energy of the whole
system as \beq \label{eqn:energypartition}
E[I+II]=E_{1}[I]+E_{2}[II]+E^{\tx{int}}[I,II] \eeq where here and
throughout the article, $I$ refers to the small region where
detailed physics are relevant, and $II$ refers to the rest of the
system. $E_{1}[I]$ represents the energy of region $I$ with region
$II$ providing appropriate boundary conditions, $E_{2}[II]$
represents the energy of region $II$ in the same sense, and
$E[I+II]$ represents the total energy of the combined system. Eq.
(\ref{eqn:energypartition}) expresses the total energy of the
system as the energy of region $I$ evaluated with the accurate
simulation method $1$, plus the energy of region $II$ evaluated
with the faster simulation method $2$, plus an energy of
interaction between the two subsystems. Typically, the crux of a
multiscale method lies in its handling of $E^{\tx{int}}$. Although
tautological, Eq. (\ref{eqn:energypartition}) can be rearranged to
yield an expression for the interaction energy: \beq
\label{eqn:tautology}
E^{\tx{int}}[I,II]=E[I+II]-E_{1}[I]-E_{2}[II] \eeq This expression
contains no new content, and merely serves to define
$E^{\tx{int}}[I,II]$, but nevertheless provides direction towards
its calculation.

The QM/MM methods are designed to achieve a goal similar to that
of the present method, namely the coupling of a quantum-mechanical
simulation with classical potentials, but in the context of
covalently bonded organic molecules.  In such systems, bonds are
localized and typically can be associated with two atoms at either
end.  The strategy often employed in QM/MM methods to couple
quantum mechanics to molecular mechanics is as
follows\cite{CuiEKFK01}: the system is divided into QM and MM
regions with a boundary that cuts across covalent bonds; $E_{QM}$
is evaluated for the QM subsystem, plus additional `link atoms'
placed on the MM side of the severed covalent bonds to mimic the
system removed from the QM region; $E_{MM}$ is evaluated for the
MM subsystem without the link atoms; and $E_{QM/MM}$ consists of
energy terms such as bond-bending terms that are left out of
$E_{QM}+E_{MM}$.  A similar methodology was developed by Broughton
et al.\cite{BroughtonABK99} for quantum-classical coupling in
silicon, the prototypical covalently-bonded bulk material.  Such
approaches rely on a somewhat artificial partitioning of the total
energy (e.g. into bond-bending and bond-stretching terms), and
hence lack a definition that could be readily generalized.  But
due to the locality of physics in covalently-bonded systems for
which QM/MM methods are appropriate, errors introduced at the
QM/MM boundary typically do not manifest themselves throughout the
system.

In metallic systems, however, the situation is quite different.
Bonds are not localized or associated with a distinct pair of
atoms.  The embedded-atom picture\cite{Daw89,DawB84} provides a
more apt description of the situation.  In the embedded atom
picture of a simple metallic system, the density of the system is
approximately the sum of charge densities of isolated atoms, and
the energetic contribution of an individual atom to the system
energy is approximately the embedding energy of the atom in a
homogeneous electron gas.  This picture, in various
forms\cite{DawB84, ErcolessiPT88, JacobsenNP87}, has been used to
great effect to create classical pair functionals for metals.  The
success of these potentials in capturing the energetics of simple
metallic systems, combined with their foundation on density
functional theory arguments, make them ideal candidates for
evaluating $E_{2}[II]$ in the present formalism.

The notions of the embedded-atom method can be extended to describe the
energetics of a metallic region (region $I$) within another metallic
region (region $II$); region $I$ is embedded within region $II$.  The exact
nature of the embedding can be formally written in the manner of Eq.
(\ref{eqn:energypartition}) with density functional
theory arguments.  We first decide which ions will be associated with
region $I$ and which will be in region $II$, and we will denote those sets of
nuclear coordinates by $\VR^{I}$ and $\VR^{II}$.  We denote the set of
all nuclei with $\VR^{\tx{tot}} \equiv \VR^I \cup \VR^{II}$.
According to the Hohenberg-Kohn theorem, the total
system energy, within the Born-Oppenheimer approximation, is given by
minimizing a functional of the total charge density:
\beq
\label{eqn:hohenener}
E[\VR^{\tx{tot}} ]=\min_{\rho} E_{\tx{DFT}}[\rho^{\tx{tot}}; \VR^{\tx{tot}} ]
\eeq
In order to be explicit, by
$E_{\tx{DFT}}[\rho;\{ \VR \}]$ we mean:
\be
\lefteqn{E_{\tx{DFT}}[\rho; \VR ]\equiv T_{\tx{s}}[\rho] + E_{\tx{H}}[\rho]
+ E_{\tx{xc}}[\rho]} \nonumber \\
&& + \sum_i \int \rho(\Vr) V_{\tx{psp}}(\Vr-\VR_i) \tx{d}\Vr + \sum_{i<j} \frac{Z_i Z_j}
{|\VR_i - \VR_j|}
\ee
where $T_{\tx{s}}$ is the non-interacting kinetic energy functional,
$E_{\tx{H}}$ is
the Hartree energy, $E_{\tx{xc}}$ is the exchange-correlation energy, and
$V_{\tx{psp}}$
is the ionic pseudopotential.  Thus $E_{\tx{DFT}}$ represents the combined
electronic and ion-ion (Madelung) energy.

If $\rho^{\tx{tot}}$ is partitioned into two sub-densities, $\rho^{I}$ and
$\rho^{II}$, such that $\rho^{\tx{tot}} = \rho^{I} + \rho^{II}$, then
the $E_{\tx{DFT}}$ can be partitioned:
\be
\lefteqn{E_{\tx{DFT}}[\rho^{\tx{tot}}; \VR^{\tx{tot}}] =
E_{\tx{DFT}}[\rho^{I}; \VR^{I} ] }&{}&{}  \nonumber \\
\label{eqn:DFTpartition} && + E_{\tx{DFT}}[\rho^{II}; \VR^{II}] +
E^{\tx{int}}[\rho^{I},\rho^{II};\VR^{I},\VR^{II}] \ee where
$E^{\tx{int}}$ is defined as in Eq. (\ref{eqn:tautology}). By
varying the total energy with respect to $\rho^{I}$, we find that
the potential felt by $\rho^{I}$ is equal to the sum of the
potential from region $I$ alone, plus an embedding potential
$V_{\tx{emb}}(\Vr)$ that completely represents the effect of
region $II$ upon region $I$: \be \frac{\delta
E_{\tx{DFT}}[\rho^{\tx{tot}};\VR^{\tx{tot}}]}{\delta \rho^{I}}
 & = & \frac{\delta E_{\tx{DFT}}[\rho^{I};\VR^{I}]}{\delta \rho^{I}} +
V_{\tx{emb}}(\Vr) \nonumber \\
\label{eqn:DFTpartition2}
V_{\tx{emb}}(\Vr) & \equiv & \frac{\delta E^{\tx{int}}[\rho^{I},
\rho^{II};\VR^{I},\VR^{II}]}{\delta \rho^{I}}
\ee
By using different approximations for the terms in Eq.
(\ref{eqn:DFTpartition2}), different coupled methods are obtained.
Wesolowski and Warshel\cite{WesolowskiW93}, building on the formalism of
Cortona
\cite{Cortona91}, used this partitioning to describe an efficient DFT method.
In their scheme $E[I]$ and $E[II]$ are treated with Kohn-Sham DFT, but
$E^{\tx{int}}$
is evaluated with ``orbital-free'' density functional theory (OF-DFT), i.e.
pure density functional theory in which the Kohn-Sham orbitals are not used
and the non-interacting kinetic energy is approximated
with an explicit functional of the density\cite{ChaconAT85,GarciaGonzalezAC96,
WangT92,WangGC98,WangGC99}.
This allows $E[I]$ and $E[II]$ to be alternately minimized in the
embedding potential of the other.
Govind et al.\cite{GovindWdC98} utilized the partitioning of Eq.
(\ref{eqn:DFTpartition}) to obtain a quantum chemistry(QC)/DFT coupled method.
There $E_1[I]$ was calculated with QC, $E_2[II]$ with DFT, and
again $E^{\tx{int}}$
was based on OF-DFT.  They used this method to explore the electronic structure
of molecules adsorbed on metal surfaces. Recently Kl\"uner et
al.\cite{KlunerGWC02} have
extended this formalism to treat adsorbed molecules in their excited
state.

\section{Formalism}\label{sec:formalism}

The present method follows in the same vein as the last few examples, to
achieve a DFT/classical  potential coupling.  The general idea of the present
methods is as follows.  $E_1[I]$ is to be calculated with DFT.
$E_2[II]$ is calculated via a classical potential.
A choice can be made for the calculation
of $E^{\tx{int}}$, which results in distinct coupling methods, which we examine
in detail below.

\subsection{Classical interaction energy}

$E^{\tx{int}}$ can be calculated using the classical potential:
\be
E^{\tx{int}}[I,II]=E_{\tx{cl}}[I+II]-E_{\tx{cl}}[I]-E_{\tx{cl}}[II]
\ee
Although this interaction energy is intended to represent the same
DFT interaction energy that appears in Eq. (\ref{eqn:DFTpartition}),
it is not contradictory to use the classical potential to evaluate it,
since the classical potential energy, evaluated for a given ionic
configuration $\VR$, can be viewed as an approximation to the
DFT functional that has been minimized with respect to the
density; that is:
\be
\label{eqn:classicalapprox}
E_{\tx{cl}}[\VR] \simeq \min_{\rho} E_{\tx{DFT}}[\rho,\VR].
\ee
This choice of interaction energy
results in a total energy of:
\beq
\label{eqn:firstscheme}
E[\VR^{\tx{tot}}] = E_{\tx{cl}}[\VR^{\tx{tot}}] - E_{\tx{cl}}[\VR^I]
 + \min_{\rho^I} E_{\tx{DFT}}[\rho^I, \VR^I]
\eeq
In this scheme, the forces on all atoms in region
$II$ are identical to forces on corresponding atoms if the classical potential
were used for the entire system; i.e. the DFT region has no effect on these
atoms.  If the cutoff length of classical potential is
$r_c$, then atoms that lie
 within region $I$ and are farther than $r_c$ from the boundary
will experience a force entirely from $E_{\tx{DFT}}[I]$; these
atoms feel a force no different than corresponding atoms in a DFT
calculation of region $I$.  The force on atoms in region $I$ that
are within $r_c$ of the boundary do not come entirely from
$E_{\tx{DFT}}[I]$, but also have contributions from
$E_{\tx{cl}}[I+II]-E_{\tx{cl}}[I]$.  These contributions should in
principle be corrections to the surface forces experienced by
these atoms from $E_{\tx{DFT}}[I]$. Classical potentials have been
developed to mimic the energetics, forces, and geometries obtained
from DFT calculations of various configurations, including
surfaces\cite{ErcolessiA94}; such potentials should be
particularly apt for the present coupling scheme.

The implementation of this method demands nothing beyond what is
required for a DFT calculation and a classical potential
calculation.  It should be noted, however, that the DFT
calculation, $E_{\tx{DFT}}[I]$, is a non-periodic calculation, and
if OF-DFT is to be used, special considerations may need to be
made for the calculation of non-periodic systems\cite{CholyK02}.

\subsection{Quantum interaction energy}
Alternatively $E^{\tx{int}}$ can be calculated more accurately with
a quantum mechanical method.
Although we only represent region $II$ by the coordinates of the ions of
region $II$ atoms and calculate the energetics with a classical potential,
there is an implicit charge density $\rho^{II}$ associated with
$E_{\tx{cl}}[\VR^{II}]$
via Eq. (\ref{eqn:classicalapprox}).  Because of this, we can consider a
more sophisticated coupling scheme where the interaction energy is based
on density functional theory.  However, in order to compute the interaction
energy via DFT when all we know about region $II$ is an approximation of its
charge density, the traditional Kohn-Sham scheme of DFT is not suitable.
In the Kohn-Sham scheme, we start with a potential and obtain the
density and energy of electrons in this potential.  Instead, we need
a means of calculating the energy of a system of electrons given their
density.  OF-DFT allows us to do this.
Thus we can write down the interaction energy in terms of OF-DFT energy
functionals:
\beq
\label{eqn:computetogether}
E^{\tx{int}}[I,II]= E_{\tx{OF}}[I+II]-E_{\tx{OF}}[I] - E_{\tx{OF}}[II]
\eeq

At first glance this seems like a useless scheme, because if DFT
is used to calculate $E^{\tx{int}}[I,II]$, we may as well use DFT
to calculate $E[I+II]$, and thus no computational expense is saved
with the coupled method. But because of the nature of many of the
useful OF-DFT functionals, this turns out not to be the case. If
$E^{\tx{int}}[I,II]$ is calculated with OF-DFT, for typical
approximate kinetic energy functionals the computation in Eq.
(\ref{eqn:computetogether}) will require a computation time that
is on the order of the computation time required to compute
$E_{\tx{OF}}[I]$, the small subsystem, rather than the time
required to compute $E_{\tx{OF}}[I+II]$.  This is because
significant cancellation is implicit in
$E_{\tx{OF}}[I+II]-E_{\tx{OF}}[II]$.

The existing approximate kinetic energy functionals differ in
accuracy and computational efficiency.  Moreover, different
choices of functional can be made for the evaluation of $E[I]$ and
$E^{\tx{int}}$, which further increases the number of possible
coupling methods.  This possibility is important because the
degree to which the computation of $E^{\tx{int}}$ can be made
efficient depends on the choice of kinetic energy functional and
the functionals that will most efficiently treat $E^{\tx{int}}$
are not necessarily accurate enough to treat the interactions
within $E[I]$.

Regardless of the choice of kinetic energy functional, the evaluation of
$E^{\tx{int}}[I,II]$ within this coupling scheme requires knowing the
electronic density of region $II$,
$\rho^{II}(\Vr)$.  In the present method, $\rho^{II}(\Vr)$ is approximated as
the sum of atomic charge densities $\rho^{\tx{at}}(\Vr)$ centered at the
region $II$ nuclei:
\be
\label{eqn:sumatomdens}
\rho^{II}(\Vr) = \sum_i \rho^{\tx{at}}(\Vr - \VR^{II}_i)
\ee
This approximation is supported by the embedded-atom picture of simple
metallic systems.  In principle, $\rho^{\tx{at}}(\Vr)$ could be a
non-spherically
symmetric density.  For example, if the arrangement of the region $II$ atoms
is always
close to the bulk lattice arrangement, then a non-spherically-symmetric charge
density that reproduces the bulk charge density when periodically tiled may
be more appropriate.  However in this article $\rho^{\tx{at}}(\Vr)$ is always
taken to be spherically symmetric.

The density in region $II$ is never explicitly represented in the
calculation, but
is given a precise form via Eq. (\ref{eqn:sumatomdens}).  Thus region $II$
is entirely described by the ionic coordinates $\VR^{II}$, and
$\rho^{II}$, the form of
which is needed to evaluate $E^{\tx{int}}$, is implicitly determined by
$\VR^{II}$.

The second coupling method is summarized by the expression for the energy
as a function of nuclear coordinates within the method:
\be
E[\VR^{\tx{tot}}] = E_{\tx{cl}}[\VR^{II}]+\min_{\rho^I} {\bigg [}
E_{\tx{OF}}[\rho^{\tx{tot}};\VR^{\tx{tot}}] \nonumber \\
 \mbox{} - E_{\tx{OF}}[\rho^{II};\VR^{II}]- E_{\tx{OF}}[\rho^I;\VR^I]
+ E_{\tx{DFT}}[\rho^I; \VR^I]{\bigg ]} \nonumber \\
\label{eqn:secondscheme} {} \ee The last term,
$E_{\tx{DFT}}[\rho^I;\VR^I]$, is written as such (and not as
$E_{\tx{OF}}[I]$) to emphasize that we could choose to compute it
either with a Kohn-Sham-type calculation or with OF-DFT, but
utilizing a more accurate kinetic energy functional than the other
OF-DFT terms.  This would allow for three distinct levels of
accuracy in the calculation: Kohn-Sham accuracy within region $I$,
OF-DFT accuracy for the coupling between regions $I$ and $II$, and
the accuracy of the classical potential in region $II$.  In this
case, $\rho^I$ would consist of a set of Kohn-Sham orbitals,
$\rho^I(\Vr)=\sum_i |\psi_i(\Vr)|^2$, and we would minimize over
the $\psi_i$: \be E[\VR^{\tx{tot}}] &=&
E_{\tx{cl}}[\VR^{II}]+\min_{\psi_i} {\bigg [}
E_{\tx{OF}}[\rho^{\tx{tot}};\VR^{\tx{tot}}]
\nonumber \\
 \mbox{} - E_{\tx{OF}}[\rho^{II};\VR^{II}] &-& E_{\tx{OF}}
[\rho^I;\VR^I]
+ E_{\tx{KS}}[\psi_i; \VR^I]{\bigg ]}
\label{eqn:secondschemeks} \\
\rho^{\tx{tot}} &=& \sum\nolimits_i |\psi_i|^2+\rho^{II}, \\
\rho^I &=& \sum\nolimits_i |\psi_i|^2
\ee
However, this interesting possibility
is not explored presently; instead we use the same type of OF-DFT calculation
for the
last four terms of Eq. (\ref{eqn:secondscheme}).  It should be noted that in
this case the last two terms
cancel, and then the total energy is given by:
\be
\lefteqn{E[\VR^{\tx{tot}}] = E_{\tx{cl}}[\VR^{II}]+}&{}&\hspace{0.8\linewidth}
 \nonumber \\
\min_{\rho^I} {\bigg [}
&&E_{\tx{OF}}[\rho^{\tx{tot}};\VR^{\tx{tot}}]
 \mbox{} - E_{\tx{OF}}[\rho^{II};\VR^{II}] {\bigg ]}
\label{eqn:OFcancel}
\ee

\subsection{Orbital-free DFT and approximate kinetic energy functionals}

Orbital-free DFT is a necessary part of the second coupling method, because
the electronic structure of region $II$ is represented only in terms of its
density via Eq. (\ref{eqn:sumatomdens}); thus in order to utilize that
information, $E^{\tx{int}}$ must be based only on the charge density and the
ionic coordinates.  Here we describe some key ideas of OF-DFT.

Hohenberg and Kohn\cite{HohenbergK64} showed that the ground state energy of a
system of electrons moving in an external potential is given by minimizing a
density functional.  Kohn and Sham\cite{KohnS65} wrote a useful
partitioning of this energy functional:
\be
\label{eqn:kohnsham}
E[\rho]=T_{s}[\rho]+E_{\tx{H}}[\rho]+ E_{\tx{xc}}[\rho]+\int V_{\tx{e-i}}(\Vr)
 \rho(\Vr) \tx{d}\Vr
\ee
where $T_{\tx{s}}$ is the non-interacting kinetic energy functional, $E_{\tx{H}}$ is
the Hartree energy, $E_{\tx{xc}}$ is the exchange-correlation energy, and
$V_{\tx{e-i}}$ is the ionic potential.
By introducing a set of fictitious non-interacting particles, we can obtain a
set of single-particle equations, the Kohn-Sham equations, that allow for
the evaluation of $E[\rho]$ with an approximate $E_{\tx{xc}}$.  The Kohn-Sham
method results in an exact evaluation of $T_{\tx{s}}[\rho_{0}]$ for the
density $\rho_{0}$ that minimizes $E[\rho]$, but the method does not provide
a means of evaluating $T_{s}[\rho]$ for an arbitrary density $\rho$.

The Kohn-Sham partitioning of the energy, Eq. (\ref{eqn:kohnsham}), has
turned out to be useful beyond the Kohn-Sham method.  Because a number of
limits of the exact non-interacting kinetic energy functional $T_{\tx{s}}[\rho]$
are known\cite{WangBook00}, $T_{\tx{s}}[\rho]$ has been
approximated by explicit density functionals constructed to satisfy one or
more of these known limits.
The orbital-free DFT methods are based on minimizing $E[\rho]$ with
$T_{\tx{s}}$ replaced with an approximate kinetic energy functional.

OF-DFT methods are typically more computationally efficient than
the Kohn-Sham method.  If the approximate $T_{\tx{s}}$ can be
evaluated with an amount of computation that scales linearly with
the system size, usually denoted by the total number of atoms $N$,
then minimizing $E_{\tx{OF}}[\rho]$ will require an amount of
computation linear in the system size (O($N$) method). Since
within OF-DFT all terms of the energy are explicit functionals of
the density, there is no need for fictitious orbitals, and the
density $\rho(\Vr)$ is the only represented variable. Thus, there
is no need to solve the single-particle Scrh\"{o}dinger equations
for the fictitious particles while maintaining their
orthogonality, operations which typically require most of the
computational effort in the Kohn-Sham approach and scale as a high
power of the system size (O($N^3$) or higher). Moreover, with the
density $\rho(\Vr)$ as the only quantity of interest in the
system, the OF-DFT methods use less memory than the Kohn-Sham
method, since the latter requires the storage and update of a
number of fictitious orbitals proportional to the system size,
each of which consumes twice the storage (as complex quantities)
needed for the density alone.

In addition to computational advantages, unlike the Kohn-Sham
method, the total energy functional $E_{\tx{OF}}[\rho]$ can be
evaluated for a given $\rho(\Vr)$. This property makes OF-DFT a
suitable candidate for computing $E^{\tx{int}}[\rho^I,\rho^{II}]$
in the second coupling method discussed in the previous
subsection.

The number of available approximate kinetic energy functionals is
sizeable, and the choice of functional is made based on considerations
of efficiency and the types of systems to be treated.  Because the
systems to be considered are simple metals with free-electron-like
charge densities, an important property that should be included in the
approximate kinetic energy functional is the correct linear response around
uniform densities:
\be
\label{eqn:linresponse}
\FT \left[\left. \frac{\delta^2 T_{\tx{s}}}{\delta \rho(\Vr)
   \delta \rho(\Vr')}\right|_{\rho_0} \right] = -\frac{1}
{\chi_{\tx{Lind}}(k)}
\ee
where $\FT$ is the Fourier transform, and $\chi_{\tx{Lind}}(k)$ is the Lindhard
response function:
\be
\chi_{\tx{Lind}}(k) = -\frac{k_F}{\pi^2}
\left[ \frac{1}{2} + \frac{1-x^2}{4x}\ln \left|\frac{1+x}{1-x}\right|
\right]
\ee
with $k_F=(3 \pi^2 \rho_{0})^{1/3}$ and $x=k/2k_F$.

A significant number of efficient functionals have been developed that
satisfy the linear response limit for a particular chosen average
density\cite{ChaconAT85,GarciaGonzalezAC96,WangT92,WangGC98,WangGC99}.  Such
functionals often consist of several
terms that are local or localized functionals, such as the Thomas-Fermi
energy and the von Weizs\"acker functionals, plus one or more  convolution
terms:
\be
\label{eqn:convolutionterm}
  T_{K}[\rho]=\int f(\rho(\Vr)) K(|\Vr-\Vr'|) g(\rho(\Vr')) \tx{d}\Vr \tx{d}\Vr'
\ee
By choosing the kernel $K(r)$ properly, the approximate functional
can be made to satisfy the correct linear response, Eq.
(\ref{eqn:linresponse}), around some chosen uniform density $\rho_{0}$.
Numerical tests indicate that among the current available efficient kinetic
energy functionals, the ones of this form are most suitable for simple
metallic systems.

However, kinetic energy functionals that contain a convolution part with
a long-ranged kernel make the efficient evaluation of $E^{\tx{int}}[I,II]$ more
difficult; the consequences of this will be discussed in the following
section.

\section{Implementation of coupling}\label{sec:implementation}

\subsection{Classical interaction energy}

The calculation of the energetics and ionic forces within the
first coupling scheme described above involve only DFT
calculations and classical potential calculations.  However, if an
ionic relaxation is to be done on the whole system, there are
several possible techniques, the optimal choice depending on the
system being relaxed.

If the partitioning of the system into regions $I$ and $II$ is
such that the time required to calculate $E_{\tx{DFT}}[I]$ is
comparable with the computation time of $E_{\tx{cl}}[I+II]$, then
ionic relaxation of the total system may be done by using a
gradient-based minimizer such as conjugate gradients methods or
quasi-Newton methods like BFGS\cite{LiuN89}. If, on the other
hand, the system partitioning is such that the time required to
evaluate $E_{\tx{DFT}}[I]$ is considerably more than that required
for the computation of $E_{\tx{cl}}[I+II]$, as is often the case,
then an alternate relaxation scheme may be more efficient. The
total system can be relaxed by using a gradient-based minimizer on
the region $I$ system alone, while fully relaxing the region $II$
ions between each ionic update of region $I$.  Gradient-based
minimizers like BFGS are only effective if the gradients involved
are indeed gradients of an underlying object function.  It is not
immediately apparent that such is the case in this
alternate-relaxation scheme, but we can demonstrate it as follows.

The energy calculated with the first coupling scheme, as a function of all
ionic coordinates, is given in Eq. (\ref{eqn:firstscheme}).  A secondary
function that only depends on the region $I$ ionic positions can be defined as:
\be
E'[\VR^I] \equiv \min_{\VR^{II}}E[\VR^{\tx{tot}}]
\ee
The useful aspect of $E'$ is that its gradient with respect to $\VR^I_i$
can be easily
evaluated:
\be
\label{eqn:easygrad}
\frac{\partial E'}{\partial \VR^I_i} &=& \frac{\partial E[\VR^{\tx{tot}}]}
{\partial \VR^I_i} + \sum_j \frac{\partial E[\VR^{\tx{tot}}]}
{\partial \VR^{II}_j} \frac{\partial \VR^{II}_{j,\tx{min}}}
{\partial \VR^I_i} \\
\label{eqn:easygrad2}
&=& \frac{\partial E[\VR^{\tx{tot}}]}{\partial \VR^I_i}
\ee
where the second term on the right of Eq. (\ref{eqn:easygrad}) vanishes
because all derivatives are evaluated at the minimum of $E[\VR^{\tx{tot}}]$
with respect
to $\VR^{II}$.  This result is analogous to the Hellmann-Feynman
theorem\cite{Hellmann37}.  The introduction of the $E'$ function allows for the
following relaxation algorithm:
\begin{itemize}
\item minimize $E[\VR^{\tx{tot}}]$ with respect to
$\VR^{II}$ while holding $\VR^I$ fixed. This only involves the
classical potential, $E_{\tx{cl}}[\VR^{\tx{tot}}]$.
\item Calculate $\min_{\rho^I} E_{\tx{DFT}}[\rho^I; \VR^I]$
and $E_{\tx{cl}}[\VR^I]$, and the forces on the region $I$ ions.  Using Eq.
(\ref{eqn:easygrad2}) the gradient of $E'$ is obtained.
\item Perform a step of a gradient-based minimization of $E'$.
\item Repeat until the system is relaxed.
\end{itemize}
In this manner, the number of DFT calculations performed is greatly reduced,
albeit at the expense of more classical potential calculations.

\subsection{Quantum interaction energy}

\begin{figure}
  \epsfig{file=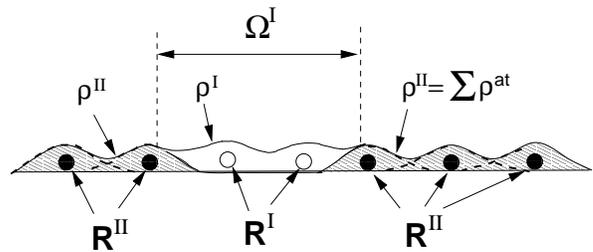, width=0.9\linewidth}
  \caption{An illustration of the partitioning of the system according to
    the coupling method with quantum interaction energy.}
  \label{fig:confine}
\end{figure}

Implementation details of the second coupling method require more
elaboration. One important point is that $\rho^I$ must be confined
to lie within a finite volume $\Omega^I$.  This region should have
significant overlap with the region where $\rho^{II}$ lies, in
order to provide coupling of the two regions. But if $\rho^I$ were
not confined to a finite volume $\Omega^I$, it could in principle
spread throughout the combined system, and during the course of
minimizing with respect to $\rho^I$ (Eq.
(\ref{eqn:secondscheme})), we would essentially be performing a
DFT calculation of the whole system. On the other hand, $\Omega^I$
should be chosen large enough so that $\rho^I$ is not artificially
confined. In the test systems we examined, we found that when
increasing the size of $\Omega^I$, a point is reached where the
results (e.g. the shape of $\rho^I$ and the forces on the ions)
change little.  The confinement of $\rho^I$ within $\Omega^I$ is
illustrated in Fig. \ref{fig:confine}.

The second coupling method maintains efficiency
due to the cancellation that occurs when $E^{\tx{int}}_{\tx{OF}}$ is computed.
Consider the computation of a local term of $E^{\tx{int}}$;
such as the exchange-correlation functional within the local density
approximation\cite{HohenbergK64} (LDA):
\be
E^{\tx{int}}_{\tx{xc}} &=& \int f_{\tx{xc}}(\rho^{\tx{tot}}) \tx{d}\Vr -
\int_{\Omega^I} f_{\tx{xc}}(\rho^I) \tx{d}\Vr
- \int f_{\tx{xc}}(\rho^{II}) \tx{d}\Vr \nonumber \\
&=& \int_{\Omega^I} \left[ f_{\tx{xc}}(\rho^{\tx{tot}}) -
f_{\tx{xc}}(\rho^I) - f_{\tx{xc}}(\rho^{II}) \right] \tx{d}\Vr
\label{eqn:Exc} \ee where $f_{\tx{xc}}(\rho) \equiv \rho
\epsilon_{\tx{xc}}(\rho)$ and we have used the fact that
$\rho^{II}(\Vr)=\rho^{\tx{tot}}(\Vr)$ for $\Vr \notin \Omega^I$.
Thus calculation of $E^{\tx{int}}_{\tx{xc}}$ is an integral over
$\Omega^I$ and not the entire system, which demonstrates our
criterion for efficiency. Any local functional of $\rho$ will
obviously be calculated efficiently in the same manner.  We note
that when the same kinetic energy functional is used for the
interaction energy and $E_{\tx{OF}}[I]$ (which is the case for the
tests performed in this paper), the cancellation exhibited in Eq.
(\ref{eqn:OFcancel}) occurs.  In this case, it is wasteful to
compute the interaction energy as in Eq. (\ref{eqn:Exc}) and then
compute and add on $E_{\tx{xc}}[\rho^I]$, as it exactly cancels
the second term of Eq. (\ref{eqn:Exc}).  Instead, we compute
directly the following quantity: \be E^{\tx{int+I}}_{\tx{xc}} =
\int_{\Omega^I} \left[ f_{\tx{xc}}(\rho^{\tx{tot}}) -
f_{\tx{xc}}(\rho^{II}) \right] \tx{d}\Vr \label{eqn:ExcintI} \ee
Similar considerations apply to the other parts of the energy
which are simple functionals of the density.  The only term that
does not fall in this category is the interaction kinetic energy
$T^{\tx{int}}_{\tx{s}}$, when it involves more sophisticated
functionals with convolution terms such as Eq.
(\ref{eqn:convolutionterm}).  For this case, we have developed an
appropriate efficient methodology, the derivation of which is
contained in appendix \ref{app:tKint}.

Particular attention must be paid to the non-local terms of
$E^{\tx{int}}$. As usual, cancellation occurs between
electron-electron, electron-ion, and ion-ion terms that eliminates
long-ranged interactions. The Hartree interaction energy is given
by: \be E^{\tx{int}}_{\tx{H}} &=& \frac{1}{2} \int
\frac{\rho^{\tx{tot}}(\Vr)\rho^{\tx{tot}}(\Vr') -
\rho^I(\Vr)\rho^I(\Vr') -
\rho^{II}(\Vr)\rho^{II}(\Vr')}{|\Vr-\Vr'|} \tx{d}\Vr
\tx{d}\Vr' \nonumber \\
&=& \int \frac{\rho^I(\Vr) \rho^{II}(\Vr')}{|\Vr-\Vr'|} \tx{d}\Vr \tx{d}\Vr' \nonumber \\
&=& \int_{\Omega^I} \rho^I(\Vr) \sum_{i} V_{\tx{H}}^{\tx{at}}(\Vr - \VR^{II}_i) \tx{d}\Vr
\ee
where
\be
V_{\tx{H}}^{\tx{at}}(\Vr) \equiv \int \frac{\rho^{\tx{at}}(\Vr')}{|\Vr-\Vr'|}\tx{d}\Vr'
\ee
Similarly the electron-ion interaction energy $E^{\tx{int}}_{\tx{e-i}}$ reduces to:
\be
\lefteqn{E^{\tx{int}}_{\tx{e-i}} = \int_{\Omega^I} \rho^I(\Vr) \sum_{i}
V_{\tx{psp}}(\Vr -\VR^{II}_i)
 \tx{d}\Vr} \nonumber \\
&&+ \int \sum_j \rho^{\tx{at}}(\Vr - \VR^{II}_j) \sum_{i} V_{\tx{psp}}
(\Vr -\VR^I_i) \tx{d}\Vr
\ee
where $V_{\tx{psp}}(r)$ is the pseudopotential representing the ion, and we
have used Eq. (\ref{eqn:sumatomdens}) to express $\rho^{II}(\Vr)$ as a sum of
$\rho^{\tx{at}}$.
Finally the ion-ion interaction energy is given by:
\be
E^{\tx{int}}_{\tx{i-i}}=\sum_{i,j} \frac{Z_i Z_j}{|\VR^I_i - \VR^{II}_j|}
\ee
The combination of all three Coulomb terms can be expressed as:
\be
E^{\tx{int}}_{\tx{H}}+E^{\tx{int}}_{\tx{e-i}}+E^{\tx{int}}_{\tx{i-i}} &=&
\int_{\Omega^I} \rho^I(\Vr) \left[
\sum_i   V^{\tx{at}}_{\tx{elec}}(\Vr-\VR^{II}_i)\right] \tx{d}\Vr \nonumber \\
&+& \sum_{i,j} \phi_{ij}(\VR_i^I - \VR_j^{II})
\label{eqn:combinedCoulomb}
\ee
where we have defined:
\be
V^{\tx{at}}_{\tx{elec}}(\Vr) &\equiv& V^{\tx{at}}_{\tx{H}}(\Vr) + V_{\tx{psp}}(\Vr), \nonumber \\
\phi_{ij}(\VR^I_i-\VR^{II}_j) &\equiv& \frac{Z_i Z_j}{|\VR^I_i - \VR^{II}_j|}
\nonumber \\
 &+ &\int V_{\tx{psp}}(\Vr - \VR^I_i) \rho^{\tx{at}}
(\Vr - \VR^{II}_j) \tx{d}\Vr
\ee
Both $V^{\tx{at}}_{\tx{elec}}(\Vr)$ and
$\phi_{ij}(\VR)$ are short-ranged functions in which the $1/R$ dependence
of the constituent terms cancel.

Within the second coupling method:
(1) we minimize the energy with respect to $\rho^I$, and
(2) we calculate the forces
on all of the ions and update their position.
In order to minimize the energy
with respect to $\rho^I$, the derivative
$\delta E^{\tx{int}} / \delta \rho^I(\Vr)$
needs to be calculated for $\Vr \in \Omega^I$.  This derivative can be
evaluated efficiently
for the local functionals like $E_{\tx{xc}}$:
\be
\frac{\delta E^{\tx{int}}_{\tx{xc}}}{\delta \rho^I(\Vr)}=f'_{\tx{xc}}(\rho^{\tx{tot}})
- f'_{\tx{xc}}(\rho^I)
\ee
where $f'_{\tx{xc}}=df_{\tx{xc}}/d\rho$.  For the long-ranged
Coulombic functionals, the derivative is given by:
\be
\frac{\delta}{\delta \rho^I(\Vr)} \left[E^{\tx{int}}_{\tx{H}} +
E^{\tx{int}}_{\tx{e-i}} +
E^{\tx{int}}_{\tx{i-i}} \right] = \sum_{i} V^{\tx{at}}_{\tx{elec}}(\Vr - \VR^{II}_i)
\ee
Evaluating this combined contribution to
$\delta E^{\tx{int}}/\delta \rho^I$ is a simple matter of evaluating
$V^{\tx{at}}_{\tx{elec}}$ for region $II$ ions located near the boundary with region $I$.
And so the gradient of the total energy with respect to $\rho^I$ is:
\be
\frac{\delta E}{\delta \rho^I(\Vr)} = \frac{\delta E_{\tx{DFT}}[\rho^I;\VR^I]}
{\delta \rho^I(\Vr)} + \frac{\delta E^{\tx{int}}_{\tx{xc}}}{\delta \rho^I(\Vr)}
+ \frac{\delta T^{\tx{int}}_{s}}{\delta \rho^I(\Vr)} \nonumber \\
\label{eqn:method2funcderiv}
+ \sum_{i} V^{\tx{at}}_{\tx{elec}}
(\Vr - \VR^{II}_i)
\ee

The calculation of the ionic forces proceeds differently for region $I$ and
region $II$ ions.  Calculation of the region $I$ ionic forces is facilitated
by the
Hellmann-Feynman theorem\cite{Hellmann37}.  If we denote the part of the
energy (Eq. (\ref{eqn:secondscheme})) that is minimized with respect to
$\rho^I$ by $G[\rho^I; \VR^I,\VR^{II}]$:
\be
G[\rho^I; \VR^I,\VR^{II}] \equiv E_{\tx{OF}}[I+II] - E_{\tx{OF}}[II] \nonumber \\
\mbox{} - E_{\tx{OF}}[I] + E_{\tx{DFT}}[I],
\ee
then we have, for the second coupling scheme:
\be
E[\VR^{\tx{tot}}]=E^{\tx{cl}}[\VR^{II}]+\min_{\rho^I} G[\rho^I; \VR^I,\VR^{II}]
\ee
and when forces on region $I$ ions are computed, the expression simplifies:
\be
\label{eqn:region1forces}
\frac{\partial E[\VR^{\tx{tot}}]}{\partial \VR^I_i} &=& \frac{\partial G}
{\partial \VR^I_i} + \int_{\Omega^I} \frac{\delta G}{\delta \rho^I(\Vr)}
\frac{\partial \rho^I_{\tx{min}}(\Vr)} {\partial \VR^I_i} \tx{d}\Vr \\
&=& \frac{\partial G}{\partial \VR^I_i}
\ee
where the last term in Eq. (\ref{eqn:region1forces}) vanishes because we have
minimized $G$ with respect to $\rho^I$, and so
$\delta G / \delta \rho^I|_{\rho^I_\tx{min}} = 0$.  So the forces on the
region $I$ ions
are determined solely by the terms of $G$ that explicitly depend on $\VR^I$;
these terms are the electron-ion energy and the ion-ion energy.  Using
Eq. (\ref{eqn:combinedCoulomb}), the force on the $i$th region $I$ ion is
given by:
\be
\lefteqn{-\VF^I_i = \frac{\partial E[\VR^{\tx{tot}}]}{\partial \VR^I_i}}
\nonumber \\
&&= \frac{\partial}{\partial \VR^I_i} {\bigg [} E_{\tx{e-i}}[I]+E_{\tx{i-i}}[I]
+ E^{\tx{int}}_{\tx{e-i}}[I,II] + E^{\tx{int}}_{\tx{i-i}}[I,II]
{\bigg ]} \nonumber \\
&&= \frac{\partial}{\partial \VR^I_i} \left[E_{\tx{e-i}}[I]+E_{\tx{i-i}}[I]
\right] + \sum_j \nabla \phi_{ij}(\VR^I_i-\VR^{II}_j) \nonumber \\
\label{eqn:reg1forces} {} \ee Thus it can be seen from Eq.
(\ref{eqn:reg1forces}) that forces on region $I$ ions are given by
the sum of the electron-ion and ion-ion forces present in
subsystem $I$ alone, and short-ranged interactions with region
$II$ ions that are nearby region $I$.

The forces on the region $II$ ions come mostly from the classical potential,
but they have contributions from $E^{\tx{int}}[I,II]$ because $\rho^{II}$ is a
function
of $\VR^{II}_j$.  Since we have not minimized with respect to $\rho^{II}$,
there
is no Hellmann-Feynman simplification when calculating the forces on
region $II$
ions, and all terms in the interaction energy contribute.  The force on the
$j$th region $II$ ion is given by:
\be
-\VF^{II}_j &=& \frac{\partial E[\VR^{\tx{tot}}]}{\partial \VR^{II}_j}
\nonumber \\
\label{eqn:F2j}
&=& \frac{\partial E^{\tx{cl}}[\VR^{II}]}{\partial \VR^{II}_j} +
\frac{\partial E^{\tx{int}}[I,II]}{\partial \VR^{II}_j}
\ee
Local functional parts of $E^{\tx{int}}$ such as the XC energy will have
a contribution to the force given by:
\be
\lefteqn{\frac{\partial E^{\tx{int}}_{\tx{xc}}}{\partial \VR^{II}_j} =} \nonumber \\
\label{eqn:reg2localcontrib}
&&-\int_{\Omega^I}
\nabla \rho^{\tx{at}}(\Vr-\VR^{II}_j) [ f'_{\tx{xc}}(\rho^{\tx{tot}})-f'_{\tx{xc}}(\rho^{II})]\tx{d}\Vr
\ee
with analogous expressions for other local contributions that may exist
in the kinetic energy functional such as the Thomas-Fermi energy.  These local
force contributions are only non-zero for region $II$ ions with an atomic density
that extends into $\Omega^I$.  It is also worth noting that the integral in Eq.
(\ref{eqn:reg2localcontrib}) need not be carried out over all of $\Omega^I$, but
only over the intersection of $\Omega^I$ with $\rho^{\tx{at}}(\Vr-\VR^{II}_j)$.

The Coulomb contributions to the region $II$ ionic forces are given by:
\be
\lefteqn{\frac{\partial}{\partial \VR^{II}_j} \left[E^{\tx{int}}_{\tx{e-i}}
+ E^{\tx{int}}_{\tx{i-i}}
 + E^{\tx{int}}_{\tx{H}} \right] = -\sum_i \nabla \phi_{ij}(\VR^I_i - \VR^{II}_j) } \nonumber \\
\label{eqn:reg2Coulombcontrib}
&& \mbox{} - \int_{\Omega^I} \rho^I(\Vr)
\nabla V^{\tx{at}}_{\tx{elec}}(\Vr - \VR^{II}_j) \tx{d}\Vr \hspace{0.4\linewidth}
\ee
This contribution also is non-zero only for region $II$ ions near the boundary of
$\Omega^I$.

If a more sophisticated kinetic energy functional with a convolution term
like Eq. (\ref{eqn:convolutionterm}) is used in $E^{\tx{int}}$, then such a term
adds considerable complication to the calculation of the forces on region $II$
ions, but these contributions nonetheless die off as we move farther from
region $I$.
Thus within the framework of this coupling scheme, the forces on region $II$ ions
take the intuitively
satisfying form of being equal to the force that arises from the classical
potential, plus a correction force for ions near the boundary of $\Omega^I$.

If the partitioning of the system between parts $I$ and $II$ is such that
region $I$ requires much longer computation than region $II$, the second
coupling
method, like the first, allows for an efficient algorithm for ionic relaxation.
We define a different partitioning of the ions as follows: we denote by
$\VR^{I'}$ the set of region $I$ ions plus all region $II$ ions
$\VR^{II}_j$ for which the interaction force $\partial E^{\tx{int}}/\partial \VR^{II}_j$
is not negligible, and we denote by $\VR^{II'}$ the rest of the $\VR^{II}$
ions. The point is that the forces on the $\VR^{II'}$ ions only depend on the
classical potential (as seen from Eq. (\ref{eqn:F2j})), and also
that $\rho^I$ does not depend on the $\VR^{II'}$
ions (as seen from Eq. (\ref{eqn:method2funcderiv})).  Thus the same
algorithm for relaxing the system in the first coupling scheme can be used,
but with $\VR^I$ replaced with $\VR^{I'}$, and $\VR^{II}$ replaced with
$\VR^{II'}$.  That is, before each relaxation step of the $\VR^{I'}$, the
$\VR^{II'}$ are to be fully relaxed.

\section{Tests}\label{sec:tests}
\begin{figure}
  \epsfig{file=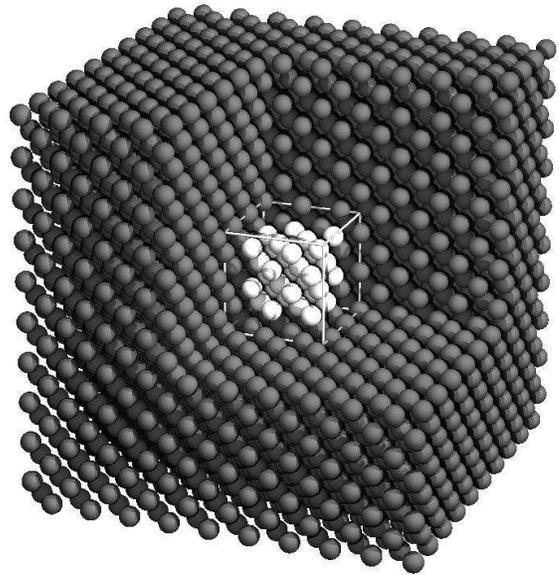, width=0.9\linewidth}
  \caption{A cut-away view of the test system.  White atoms belong to region $I$,
    and dark atoms belong to region $II$.  For the coupling method with quantum
    interaction energy, the region $\Omega^I$ is shown by the white cube.}
  \label{fig:testsystem}
\end{figure}

In order to test the present coupling methods, we have focused on
a simple coupled system that is readily analyzed.  The system
consists of $10 \times 10 \times 10$ cubic unit cells (4 atoms
each) of crystalline fcc aluminum.  The innermost $2 \times 2
\times 2$ unit cells (32 atoms total) are considered to be region
$I$, and all atoms outside are considered to be in region $II$.
Region $II$, which is treated with the classical potential, is
treated as a periodic system in order to remove effects of
surfaces from the simulation.  So in fact the test system consists
of an infinite array of 32-atom Al clusters treated quantum
mechanically, embedded in an Al bulk treated by classical
potentials. Obviously, if there is good coupling between region
$I$ and region $II$, the entire system should simply behave like
pure bulk fcc Al, making it easy to evaluate the quality of the
coupling. This test system is illustrated in Fig.
\ref{fig:testsystem}.

In all of the present tests, region $I$ is treated with OF-DFT.  However, the
particular kinetic energy functional used differs among the tests.
The Al ions are represented with the Goodwin-Needs-Heine
local pseudopotential\cite{GoodwinNH90}.
For all tests, the classical potential used is the ``glue'' potential of
Ercolessi and Adams\cite{ErcolessiPT88}, which has the embedded-atom method
(EAM) form:
\be
E[\VR_i] = \sum_i F \left[ \sum_{i \neq j} \rho^{\tx{EAM}}(|\VR_{ij}|) \right]
+ \frac{1}{2}\sum_{i \neq j} \phi(|\VR_{ij}|)
\ee
The EAM potential has been scaled both in $\Vr$ and in
energy:
\be
F[\rho] &\rightarrow& \alpha F[\rho], \nonumber \\
 \rho^{\tx{EAM}}(\VR) &\rightarrow& \rho^{\tx{EAM}}(\beta \VR), \nonumber \\
\phi(\VR) &\rightarrow& \alpha \phi(\beta \VR)
\ee
with $\alpha$ and $\beta$ chosen so that the potential yields precisely the
same lattice constant and bulk modulus of fcc Al simulated with OF-DFT
employing the particular kinetic energy functional used in that test.
This is in accord with the philosophy that the coupled
simulation should behave as if the accurate method were used for the entire
system.  But this procedure is also done so that a ``fair'' assessment of the
coupling itself can be made; we wish to examine errors in the present coupling
methods themselves and the approximations involved in them, but not the errors
coming from a trivial incompatibility between energy methods.  To make the
classical potential even more compatible with the OF-DFT method, we could
re-determine the form of the classical potential using the method that
Ercolessi and Adams originally used to develop their
potential\cite{ErcolessiA94}:
 perform a large number of reference energetic calculations of Al using OF-DFT,
and find the EAM-type potential that best reproduces these results.  This
would be a rather involved procedure, so we have chosen to simply scale the
potential.

\subsection{Test of classical interaction energy method}

In the first coupling method (Eq. (\ref{eqn:firstscheme})), the
energetics of region $I$ was treated with OF-DFT, and the kinetic
energy functional used was one developed by Wang et
al.\cite{WangGC99}, with a density-dependent kernel, and
parameters $\{\alpha, \beta, \gamma, \rho_{\ast}\}= \{5/6 +
\sqrt{5}/6, 5/6- \sqrt{5}/6, 2.7, 0.183 \text{ \AA}^{-3}\}$ (in
the notation of Ref. \cite{WangGC99}). This functional has six
convolution terms of the form of Eq. (\ref{eqn:convolutionterm}).
The classical potential was scaled to match the lattice constant
($4.033 \text{ \AA}$) and bulk modulus (55.7 GPa) of fcc Al
obtained by this kinetic energy functional.

\begin{figure}
  \epsfig{file=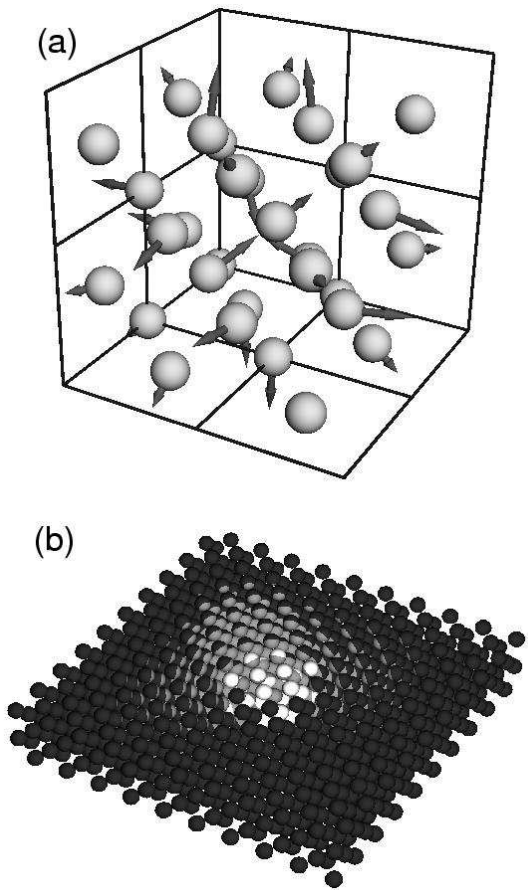, width=0.9\linewidth}
  \caption{Test of coupling method with classical interaction energy.
    (a) The forces on the region $I$ atoms when the
    atomic positions are at the perfect lattice positions.  The force factors
    are scaled so that a vector of length one lattice constant corresponds to $1$ eV/\AA.
    (b) The relaxed region $I$ and 2 atomic positions shown in white and black,
    respectively. The perfect lattice sites are drawn as gray spheres of
    a slightly smaller radius.}
  \label{fig:method1}
\end{figure}

The system was initially arranged in the perfect fcc lattice
configuration The forces on the region $II$ atoms were identically
zero, since they come entirely from $E^{\tx{cl}}[I+II]$, which is
at a minimum in the initial configuration.  However, the forces on
the region $I$ atoms are not zero, as the OF-DFT and EAM forces do
not perfectly cancel each other.  The average magnitude of the
force difference per atom between the OF-DFT and EAM calculations
of region $I$ was $0.34$ eV/\AA.  These initial forces on the
region $I$ atoms are shown in Fig. \ref{fig:method1}(a), with the
drawn force vectors scaled so that a force of $1$ eV/\AA would
extend one lattice constant.  Then the coupled system was relaxed.
After relaxation, it was found that the atomic positions deviated
from the correct fcc lattice positions by an average of $0.004$
\AA per atom.  The average deviation of just the region $I$ atoms
was $0.076$ \AA per atom.  The atomic deviation is shown in Fig.
\ref{fig:method1}(b), in which the relaxed atomic positions for
the region $I$ and region $II$ atoms are drawn as white and black
balls, respectively, and the correct lattice positions are drawn
as gray balls of a slightly smaller radius.  Note that only the
four $(100)$ layers that include region $I$ are shown. From this
diagram it can be seen that in general the relaxed atomic
positions deviate from the the perfect lattice positions by
bulging out from region $I$ slightly, with the deviation
decreasing with increasing distance from region $I$.

\subsection{Test of quantum interaction energy method}

The second coupling method was applied to the same simple test
system. The kinetic energy functional employed was again from Ref.
(\cite{WangGC99}), but in this case it was a functional with a
density-independent kernel, with parameters $\{\alpha,
\beta\}=\{5/6 \pm \sqrt{5}/6\}$.  A different functional was
chosen for this test because of its simpler form: it contains only
one convolution term of the form of Eq.
(\ref{eqn:convolutionterm}), while the functional used in the test
of the first coupling method had six. This makes the evaluation of
the kinetic interaction energy, $T_{\tx{s}}^{\tx{int}}$, simpler.
Furthermore, this functional performs well for structures that do
not deviate much from the bulk system. This functional was found
to be inapplicable to the test of the first coupling method,
because in that approach the calculation of $E_{\tx{OF}}[I]$
amounts to an isolated cluster; in that case, there is no
embedding potential from region $II$ and the minimization with
respect to $\rho^I$ does not converge. For bulk fcc Al, however,
this simpler functional employed to test the quantum interaction
energy method, performs quite well producing an equilibrium
lattice constant of $4.035$ \AA and a bulk modulus of $71.9$ GPa.

\begin{figure}
  \epsfig{file=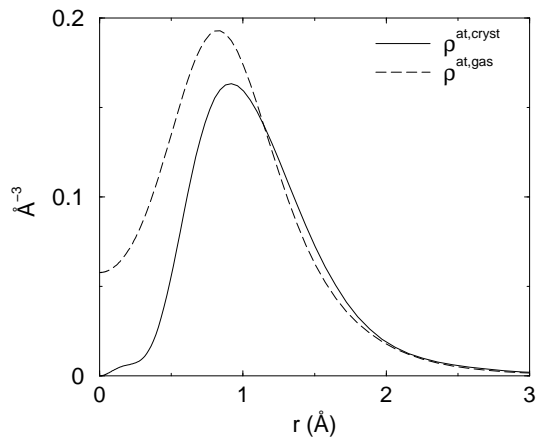, width=0.8\linewidth}
  \caption{The two atomic densities used to represent $\rho^{II}$.}
  \label{fig:rhoatomic}
\end{figure}

Another aspect of the second coupling method is the choice of atomic
density functions $\rho^{\tx{at}}$ representing $\rho^{II}$ through
Eq. (\ref{eqn:sumatomdens}).  Two different choices of $\rho^{\tx{at}}$ were
tried.  One choice, $\rho^{\tx{at,gas}}$,
was the valence density from a Kohn-Sham calculation of
an isolated Al atom, represented with the same
pseudopotential\cite{GoodwinNH90}.
The other choice, $\rho^{\tx{at,cryst}}$ was again a spherically symmetric
charge density chosen such
that the charge density that results from periodically superposing it
on an fcc lattice most closely matches the charge density coming from
an OF-DFT calculation of bulk fcc Al.  The
desired spherically symmetric charge density $\rho^{\tx{at,cryst}})(r)$
minimizes:
\be
\int_{\Omega} \left[ (\delta^{\tx{fcc}} \ast \rho^{\tx{at,cryst}})(\Vr) -
\rho^{\tx{fcc}}(\Vr)
 \right]^2 \tx{d}\Vr
\ee
where $\delta^{\tx{fcc}}(\Vr)$ is an infinite fcc lattice of delta functions,
$\ast$ denotes convolution, and $\Omega$ is one unit cell;
$\rho^{\tx{fcc}}(\Vr)$ is the valence charge density of the fcc crystal.
In reciprocal space, this becomes:
\be
\Omega \sum_{\VQ} \left| \tilde{S}_{\VQ}\tilde{\rho}^{\tx{at,cryst}}(Q) -
\tilde{\rho}^{\tx{fcc}}_{\VQ} \right|^2
\ee
where $\tilde{S}_{\VQ}$ is the structure factor. This is minimized by
requiring:
\be
\tilde{\rho}^{\tx{at,cryst}}(Q)= \frac{ \langle \tilde{\rho}^{\tx{fcc}}_{\VQ}
\rangle_{Q}}{\langle \tilde{S}_{\VQ}\rangle_{Q}}
\ee
where $\langle \cdots \rangle_{Q}$ denotes an averaging over reciprocal
lattice vectors $\VQ$ of length $Q$.  This $\tilde{\rho}^{\tx{at,cryst}}(Q)$
was then
used to construct a radial charge density $\rho^{\tx{at,cryst}}(r)$ that was
commensurate
with $\tilde{\rho}^{\tx{at,cryst}}(Q)$ at the values of $Q$ where the latter
was defined.
The two charge density choices are shown in Fig. \ref{fig:rhoatomic}.

\begin{figure*}
  \epsfig{file=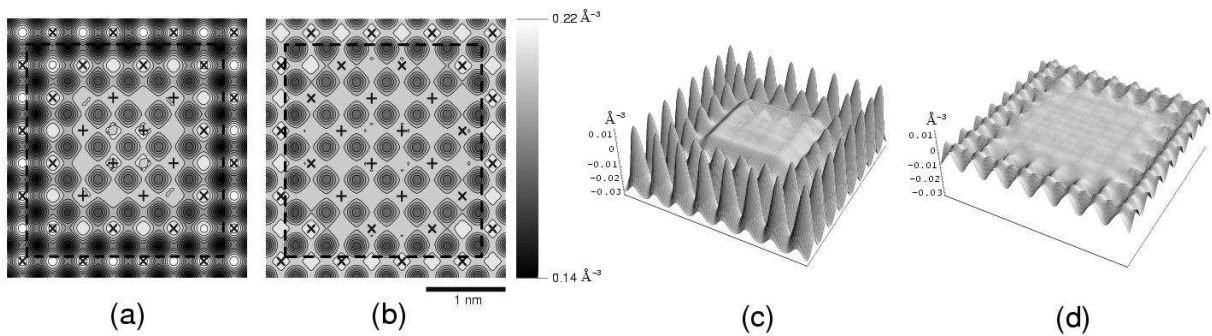, width=0.9\linewidth}
  \caption{Results for the quantum interaction energy method.  (a) and (b)
    are contour
  plots of $\rho^{\tx{tot}}$ with $\rho^{II}$ given by superpositions of
  $\rho^{\tx{at,gas}}$ and $\rho^{\tx{at,cryst}}$, respectively.  The
  boundary of $\Omega^I$ is shown with a dashed line, and the positions of the
  region $I$ atoms lying in this plane are indicated by $(+)$, and region $II$ atoms
  by $(\times)$. (c) and (d) show the difference between these two densities
  and the ``correct'' density coming from a purely OF-DFT calculation of the
  whole system.}
  \label{fig:coupling2}
\end{figure*}

The second coupling method was tested on the same system used to
test the first coupling method.  With the second method, however,
we must choose the form of the atomic density representing
$\rho^{II}$, and the extent of the region $\Omega^I$.  With
respect to the choice of $\Omega^I$, we have found the following
general behavior: if $\Omega^I$ is chosen to be too small, then
after minimization with respect to $\rho^I$, there is an excess
buildup of charge near the boundary of $\Omega^I$.  This in turn
results in a net attraction of the region $I$ ions toward the
boundary of $\Omega^I$.  This is remedied by an increase in the
size of $\Omega^I$.  When $\Omega^I$ is increased further still,
the results (the ionic forces, and $\rho^I$) are found to change
only very slightly.  We note that regardless of the size of
$\Omega^I$, the total density is always found to be continuous at
the boundary due to the high energy that the kinetic energy
functional assigns to a discontinuity in the density.  In Fig.
(\ref{fig:coupling2}) we have plotted the total charge density
after minimizing with respect to $\rho^I$, with $\rho^{II}$ given
by (a) a superposition of $\rho^{\tx{at,gas}}$, and (b) a
superposition of $\rho^{\tx{at,cryst}}$.  The particular slice of
the charge density is a $(100)$ plane that passes through one of
the central atomic planes of the region $I$ cluster.  In (c) and
(d) we have plotted the difference between these coupled charge
densities and the density of this system when computed entirely
with OF-DFT.  In general, using $\rho^{\tx{at,cryst}}$ results in
a more accurate total charge density.  From (c) and (d), it is
clear that the superposition of $\rho^{\tx{at,cryst}}$ reproduces
the pure OF-DFT crystal charge density better than
$\rho^{\tx{at,gas}}$ both for points $\Vr$ well within $\Omega^I$,
as well as at the boundary of $\Omega^I$.

It turns out, however, that the forces on the ions, for both choices of
$\rho^{\tx{at}}$, are comparable.  Also comparable is the amount of deviation
from the perfect lattice positions upon atomic relaxation for both
choices.  The exact numbers for these quantities for both coupling methods
and the two choices of $\rho^{\tx{at}}$ are summarized in Table
\ref{tab:coupling}.

\begin{table}
\caption{Summary of the performance of the two coupling methods,
and the two choices for $\rho^{\tx{at}}$ in the quantum
interaction energy method. $\VF^I_{\tx{max}}$,
$\VF^{II}_{\tx{max}}$ are the maximum forces on region $I$ atoms
and region $II$ atoms, and $d^I_{\tx{max}}$, $d^{II}_{\tx{max}}$
the maximum displacements from the perfect lattice positions upon
relaxation for region $I$ and region $II$ atoms.}
\label{tab:coupling}
\begin{tabular}[t]{lcccc}
\hline\hline
interaction   & $F^I_{\tx{max}}$ &
$F^{II}_{\tx{max}}$
  & $d^I_{\tx{max}}$ & $d^{II}_{\tx{max}}$  \\
energy &(eV/\AA) & (eV/\AA) &  (\AA) &  (\AA) \\
\hline
classical & 0.45 & 0 & 0.12 & 0.048 \\
quantum, $\rho^{\tx{at,gas}}$ & 0.12  & 0.27 & 0.12 & 0.15 \\
quantum, $\rho^{\tx{at,cryst}}$ & 0.094  & 0.37 & 0.217 & 0.265 \\
\hline\hline
\end{tabular}
\end{table}

\section{Conclusions}\label{sec:conclusions}

The coupling of classical and quantum simulation in simple metals
involves a set of challenges quite different than those for the
coupling of covalently-bonded materials and molecules, and hence
requires a different set of approaches.  We have presented here
two methods for combining classical and quantum mechanical
simulation of simple metals. Both are based on a similar
partitioning of the energy of the system, but they differ in how
the energy of interaction between the classical and quantum
mechanical parts of the system are treated. We have presented
numerical implementations that allow both coupling methods to be
efficient.

Within the first coupling method, in which the interaction energy
is determined from the classical potential, forces in the
classical region are fully determined by the classical potential.
Forces in the quantum region are determined by both classical and
quantum energetics, the quantum energetics dominating well within
the quantum region.  A major practical advantage of this approach
is that, if region $I$ contains many different atomic species
while region $II$ contains only one atom type, there is no need
for a classical potential for each species and their interactions;
if the various species of atoms are well within region $I$, then
the potential representing them does not matter at all as
interactions in this region are treated purely with quantum
mechanics.

Within the second coupling method, in which the interaction energy
is determined via OF-DFT, forces in the classical region are
determined mostly by the classical potential, with quantum
contributions to atomic forces near the boundary of the regions.
Forces in the quantum region are determined fully by quantum
energetics.  Within the quantum region, the charge density
accurately reproduces the correct charge density, and smoothly
joins with the implicit density (given by a sum of atomic
densities) of the classical region.

Test results indicate that the second coupling method yields more
accurate forces on the atoms in the quantum region than the first
method, but that the first method yields more accurate forces for
the atoms in the classical region.  This may be due, to some
extent, to the less-accurate OF-DFT method used in the test of the
second coupling method.  The first coupling method also yielded a
better relaxed structure, probably due to its better treatment of
forces on the classical atoms.  However, unlike the first method,
the second coupling method results in a more accurate charge
density within the quantum mechanical region, allowing for an
accurate treatment of physical problems such as the introduction
of impurities, where the background density is important. We also
find that a superposition of atomic charge densities can reproduce
the actual charge density well for a simple metallic system, given
an appropriate choice for the atomic charge density; this allows
for a smooth density transition at the boundary between regions.

Clearly, an important issue affecting the coupling quality for
both methods is the agreement between forces from the DFT methods;
within both methods there are atoms whose forces are determined by
a combination of quantum and classical energetics, and the more
closely the two energetics agree, the better the coupling will be.
An improvement in the quality of the coupling might be obtained if
the classical potential employed in region $II$ is optimized to
closely reproduce the DFT energetics; this is also in accord with
the multiscale philosophy that a coupled simulation should act as
if the most accurate method were used to simulate the entire
system.

\appendix
\section{Evaluating the interaction energy for complex kinetic energy
 functionals}
\label{app:tKint}

We describe here how the interaction energy can be efficiently
calculated when the approximate kinetic energy functional used is of a more
complicated form, containing a convolution term of the form of Eq.
(\ref{eqn:convolutionterm}).  That is, we will describe a method for
evaluating:
\be
T_{\tx{K}}^{\tx{int}}[\rho^I,\rho^{II}]=\int f_{12}(\Vr) K(\Vr-\Vr') g_{12}(\Vr') \tx{d}\Vr \tx{d}\Vr'
\nonumber \\
- \int f_{1}(\Vr) K(\Vr-\Vr') g_{1}(\Vr') \tx{d}\Vr \tx{d}\Vr' \nonumber \\
\label{eqn:TKint}
- \int f_{2}(\Vr) K(\Vr-\Vr') g_{2}(\Vr') \tx{d}\Vr \tx{d}\Vr',
\ee
where we have defined $f_1(\Vr) \equiv f(\rho^I(\Vr))$, $f_{12}(\Vr) \equiv
f(\rho^I(\Vr)+\rho^{II}(\Vr))$, and so on.
Then we define two new functions,
\be
F(\Vr) &\equiv& f_{12}(\Vr) - f_2(\Vr), \nonumber \\
G(\Vr) &\equiv& g_{12}(\Vr) - g_2(\Vr)
\ee
Note that $F(\Vr)$ and $G(\Vr)$ are zero for points $\Vr \notin \Omega^I$.  Using
$F$ and $G$ we can re-express Eq. (\ref{eqn:TKint}) as:
\be
T_{\tx{K}}^{\tx{int}}=\int_{\Omega^I} F (\Vr) (K \ast G)(\Vr) \tx{d}\Vr
- \int_{\Omega^I} f_1(\Vr) (K \ast g_1)(\Vr) \tx{d}\Vr \nonumber \\
 \mbox{} + \int_{\Omega^I} F(\Vr) (K \ast g_2)(\Vr) \tx{d}\Vr
+ \int_{\Omega^I} G(\Vr) (K \ast f_2)(\Vr) \tx{d}\Vr \nonumber \\
\label{eqn:TKV1}
\ee
where we have now defined:
\be
\label{eqn:Kconv}
(K \ast G)(\Vr) &\equiv& \int K(\Vr-\Vr') G(\Vr') \tx{d}\Vr', \mbox{ etc.}
\ee
We point out that if this interaction energy is being calculated in a
coupled simulation in which the energetics of region $I$ are calculated
using the same kinetic energy (i.e. $E[I]$ and $E^{\tx{int}}[I,II]$ being
treated
at the same level of theory), then the final term of Eq. (\ref{eqn:TKV1})
is equal to and will cancel with the corresponding term in $E[I]$.

So with Eq. (\ref{eqn:TKV1}) we have expressed $T_{\tx{K}}^{\tx{int}}$ purely in terms of
intergrals over $\Omega^I$; the problem is now reduced to efficiently calculating
the functions $(K \ast f_2)(\Vr)$ and $(K \ast g_2)(\Vr)$ for points
$\Vr$ within $\Omega^I$.  A
straightforward integration for each point $\Vr \in \Omega^I$ is not an option,
because $K(r)$ is typically long-ranged, and thus determining
$(K \ast f_2)(\Vr)$ at
one single point $\Vr$ would require an integration over the volume of the
whole coupled system, which would be highly inefficient.  We now describe
a method for determining $(K \ast f_2)(\Vr)$, and $(K \ast g_2)(\Vr)$ can be
determined with precisely the same method.

In earlier work\cite{CholyK02} we have developed a method for
efficiently evaluating convolutions like Eq. (\ref{eqn:Kconv}) when the
convolution kernel $K(r)$ is of the particular form typically encountered
in kinetic energy functionals involving convolution
terms\cite{ChaconAT85,GarciaGonzalezAC96, WangT92,WangGC98,WangGC99}.  We will
invoke this method to determine $(K \ast f_2)$.  In this method, the
kernel is fit in reciprocal space with the following form:
\be
\label{eqn:kspacefit}
\tilde{K}(k) &\simeq& \sum_i \tilde{K}_i(k), \nonumber \\
\tilde{K}_i(k) &=& \frac{P_i k^2}{k^2 + Q_i}
\ee
where  $P_i$, $Q_i$ are complex fitting parameters.  The kernels encountered
in many kinetic energy functionals are well-fit with this form, with
4 terms.  The kernel in real space can be expressed as the sum of the inverse
Fourier transform of each term of Eq. (\ref{eqn:kspacefit}):
\be
K(\Vr) &\simeq& \sum_i K_i(\Vr), \nonumber \\
K_i(\Vr) & \equiv & P_i \delta(\Vr) - P_i Q_i \frac{e^{-\sqrt{Q_i}r}}{4 \pi r}
\ee
Thus $(K \ast f_2)$ can be written as the sum of separate
convolutions:
\be
\label{eqn:sepconvsum}
(K \ast f_2)(\Vr) &=& \sum_i (K_i \ast f_2)(\Vr), \\
\label{eqn:separateconv}
(K_i \ast f_2)(\Vr) &\equiv& \int K_i(\Vr - \Vr') f_2(\Vr') \tx{d}\Vr'
\ee
Because in reciprocal space the $(K_i \ast f_2)$ satisfy:
\be
\left[ k^2 + Q_i \right] (\widetilde{K_i \ast f_2})(\Vk) = P_i k^2 \tilde{f}_2
(\Vk),
\ee
in real space they satisfy:
\be
\label{eqn:helm}
\left[ \nabla^2 - Q_i \right] (K_i \ast f_2)(\Vr) = P_i \nabla^2 f_2 (\Vr)
\ee
i.e. they are solutions to (complex) Helmholtz equations which can be solved
with conjugate-gradient-based methods\cite{FreundN91}; such methods are
efficient
and only involve operations within $\Omega^I$.  The solutions to Eqs.
(\ref{eqn:helm}) are only well-defined when boundary conditions for
$(K_i \ast f_2)(\Vr)$ are supplied.

\begin{figure}
  \epsfig{file=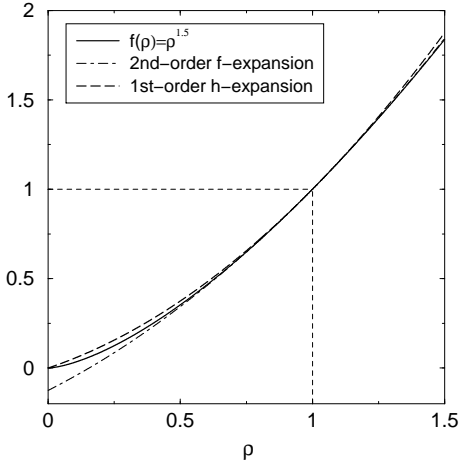, width=0.7\linewidth}
  \caption{A demonstration of the Taylor expansion of Eq.
(\ref{eqn:hexpansion}) compared to a direct Taylor expansion of
$f(\rho)=\rho^{1.5}$ about $\rho_0=1$.}
  \label{fig:hexpansion}
\end{figure}

We propose the use of Dirichlet boundary conditions.  The value of
$(K_i \ast f_2)(\Vr)$ for points $\Vr$ on the boundary of $\Omega^I$ can be found
by evaluating the convolutions, Eqs. (\ref{eqn:separateconv}).
Because of the regular nature of $\rho^{II}(\Vr)$, being the sum of atomic
densities, an efficient method for evaluating these convolutions exists.
The form of the convolution that needs to be evaluated is:
\beq
\label{eqn:convform}
(K_i \ast f_2)(\Vr)=\int K_i(\Vr-\Vr') f\left( \sum_j
\rho^{\tx{at}}(\Vr'-\VR^{II}_j)
\right)
\tx{d}\Vr'
\eeq
If $f(\rho)$ were a linear function, then this would reduce to a sum of
pair functions.  For many kinetic energy functionals, $f(\rho)$ is
not linear, but equal to some power of $\rho$: $f(\rho)=\rho^{\alpha}$.  This
leads us to consider a Taylor expansion of $f(\rho)$ about some average
density $\rho_0$.
This Taylor expansion suffers in places where $\rho^{II}(\Vr)$ is
near $0$, which occurs, for instance, in the center of $\Omega^I$.  An expansion
that is much more accurate down to small values of $\rho$ is obtained
if we Taylor expand the function $h(\rho) \equiv f(\rho)/\rho$ and express
$f(\rho^{II})$ in terms of this expansion:
\be
f \left(\sum_j \rho^{\tx{at}}(\Vr'-\VR^{II}_j) \right) \simeq \left(\sum_j
\rho^{\tx{at}}(\Vr'-\VR^{II}_j) \right) \nonumber \\
\label{eqn:hexpansion}
\times \left[ h(\rho_0) + h'(\rho_0) \left[ \sum_k \rho^{\tx{at}}(\Vr'-\VR^{II}_k)
-\rho_0 \right] + \cdots
\right]
\ee
In Fig. \ref{fig:hexpansion} we illustrate the effectiveness of
(\ref{eqn:hexpansion}) compared to expanding $f(\rho)$ directly when
$f(\rho)=\rho^{1.5}$ and $\rho_0=1$.  Although the
h-expansion is taken only to first order, while the f-expansion is taken to
second order, the h-expansion is seen to be more accurate at small $\rho$.

\begin{figure}
  \epsfig{file=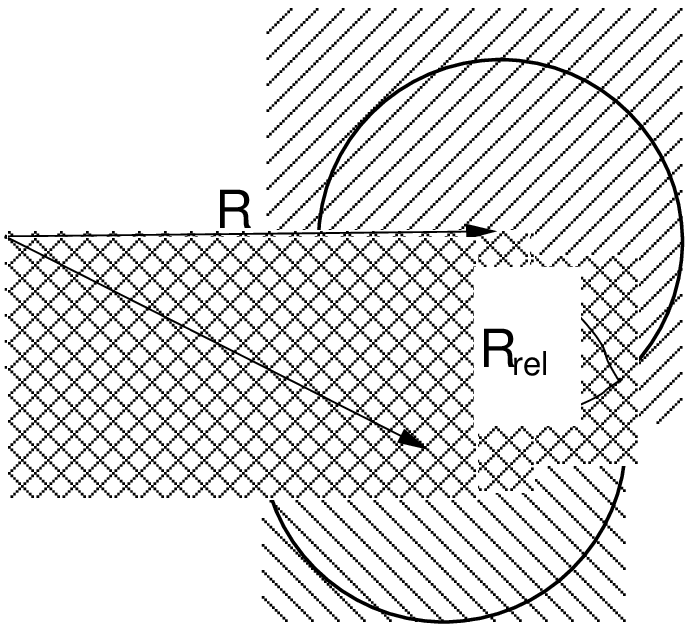, width=0.6\linewidth}
  \caption{New coordinates $\VR_{\tx{c}}$ and $\VR_{\tx{rel}}$
for evaluating the three-center integrals of $L^{(2)}_i$.}
  \label{fig:newcoords}
\end{figure}

Upon substitution of the expansion of Eq. (\ref{eqn:hexpansion}) in the
convolution, Eq. (\ref{eqn:convform}), we find:
\be
(K_i \ast f_2)(\Vr) &\simeq& \left[ h(\rho_0) -
\rho_0 h'(\rho_0) \right] \sum_j L^{(1)}_i(\Vr - \VR^{II}_j)
 \nonumber \\
&& \mbox{} + h'(\rho_0) \sum_{j,k} L^{(2)}_i(\Vr-\VR^{II}_j,\Vr-\VR^{II}_k),
\nonumber \\
L^{(1)}_i(\VR) &\equiv& \int K_i(\Vr')\rho^{\tx{at}} (\VR-\Vr')\tx{d}\Vr', \nonumber \\
L^{(2)}_i(\VR,\VR') &\equiv&  \int K_i(\Vr') \rho^{\tx{at}}(\VR -\Vr')
\rho^{\tx{at}}(\VR'-\Vr') \tx{d}\Vr' \nonumber \\
{}
\ee
$L^{(1)}_i$ is the convolution of an atomic density with $K_i(r)$.
$L^{(2)}_i(\VR,\VR')$ is the convolution of the product of two
atomic densities with $K(r)$, and consequently vanishes when the two atomic
densities do not overlap.  The integrand is non-zero only where the overlap
occurs.  It thus makes sense to express $L^{(2)}_i$ in
terms of new coordinates, illustrated in Fig. \ref{fig:newcoords}:
\be
M_i(\VR_{\tx{c}},\VR_{\tx{rel}}) \equiv L^{(2)}_i(\VR_{\tx{c}} -
\smhalf \VR_{\tx{rel}}, \VR_{\tx{c}} + \smhalf \VR_{\tx{rel}})
\ee
In the case of a spherically symmetric
$\rho^{\tx{at}}(\Vr)$, $L^{(1)}_i(\VR)$ is a function only of $|\VR|$, and
$M_i(\VR_{\tx{c}}, \VR_{\tx{rel}})$ will depend only on $|\VR_{\tx{c}}|$,
$|\VR_{\tx{rel}}|$, and
$\VR_{\tx{c}} \cdot \VR_{\tx{rel}}$.  We now argue that the dependence on
$\VR_{\tx{c}} \cdot \VR_{\tx{rel}}$ is weak.  We can write the expression for
$M_i$ as:
\be
M_i(\VR_{\tx{c}},\VR_{\tx{rel}})=
\int K_i(\VR_{\tx{c}} - \Vr) \rho^{\tx{at}}(\Vr- \smhalf \VR_{\tx{rel}})
\nonumber \\
\label{eqn:Mform}
\mbox{} \times \rho^{\tx{at}}(\Vr + \smhalf \VR_{\tx{rel}}) \tx{d}\Vr
\ee
If we expand $K$ about $\VR_{\tx{c}}$, we find:
\be
\lefteqn{M_i(\VR_{\tx{c}},\VR_{\tx{rel}}) = \int \left[K_i(|\VR_{\tx{c}}|) -
\Vr \cdot
\nabla K_i(\VR_{\tx{c}}) + \cdots \right]} &&\hspace{0.1\linewidth}
\nonumber \\
& \mbox{} \times \rho^{\tx{at}}(\Vr- \smhalf \VR_{\tx{rel}}) \rho^{\tx{at}}
(\Vr + \smhalf
\VR_{\tx{rel}}) \tx{d}\Vr &
\ee
The integrals over the odd powers of $\Vr$ in the expansion of $K_i$ vanish
by symmetry.  Thus if we truncate the series at first order, the first order
term vanishes, leaving only the $0$th order term:
\be
\label{eqn:approxM}
M_i(\VR_{\tx{c}},\VR_{\tx{rel}}) \simeq K_i(|\VR_{\tx{c}}|)
P(|\VR_{\tx{rel}}|), \\
P(|\VR_{\tx{rel}}|) \equiv \int \rho^{\tx{at}}(\Vr)\rho^{\tx{at}}
(\Vr-\VR_{\tx{rel}})\tx{d}\Vr
\ee
Truncating the expansion of $K_i$ at the first order is reasonable, because
$K_i(r)$ oscillates around the Fermi wavelength of the system, a length
scale close to that of $\rho^{\tx{at}}$.  This approximate form,
Eq. (\ref{eqn:approxM}) will behave quite badly at small $R_{\tx{c}}$, because
$K_i$ diverges at the origin, and the radial averaging of this divergence
that occurs in Eq. (\ref{eqn:Mform}) is not reflected in Eq.
(\ref{eqn:approxM}).  Thus we replace $K_i(|\VR_{\tx{c}}|)$ there with the
convolution of $K_i$ with a Gaussian of unit weight and a variance $r_0$
given roughly by the length-scale of the overlap regions of the atomic
densities (i.e. some fraction of the range of $\rho^{\tx{at}}(\Vr)$):
\be
M_i(\VR_{\tx{c}},\VR_{\tx{rel}}) &\simeq& K'_i(|\VR_{\tx{c}}|)
P(|\VR_{\tx{rel}}|), \nonumber \\
K'_i(\Vr) &\equiv& (K_i \ast w)(\Vr), \nonumber \\
w(\Vr) &\equiv& \pi^{-3/2} r_0^{-3} e^{-(r/r_0)^2}
\ee

Summarizing these results, we find that we can approximately evaluate
$(K_i \ast f_2)$ as:
\be
\lefteqn{(K_i \ast f_2)(\Vr) \simeq \left[h(\rho_0)-\rho_0 h'(\rho_0) \right]
 \sum_j L^{(1)}_i(\Vr - \VR^{II}_j)}
 \nonumber \\
\label{eqn:fastkf}
&&\mbox{} + h'(\rho_0) \sum_{<j,k>} K'_i \left( \left| \Vr -
\smhalf(\VR^{II}_j+\VR^{II}_k)
\right| \right) P \left(\left| \VR^{II}_j - \VR^{II}_k \right| \right) \nonumber \\
{}
\ee
where the summation over $<j,k>$ indicates that we need only sum over
pairs of region $II$ atoms with overlapping densities.  The derivation of
Eq. (\ref{eqn:fastkf}) involved several approximations, and thus is not
expected to be precise.  We only propose that Eq. (\ref{eqn:fastkf})
be used to generate the boundary conditions for Eqs. (\ref{eqn:helm}) that
determine the $(K_i \ast f_2)(\Vr)$ within region $\Omega^I$, and we have found
that the
resulting $(K_i \ast f_2)(\Vr)$ is more dependent on the source term than the
boundary conditions.  Nevertheless, because of the inaccuracies of Eq.
(\ref{eqn:fastkf}), we define a new region, $\Omega^{I'}$, that contains and extends
a bit beyond $\Omega^I$, and we use Eq. (\ref{eqn:fastkf}) to obtain the boundary
conditions for points $\Vr$ that lie on the boundary of $\Omega^{I'}$, and we solve
Eqs. (\ref{eqn:helm}) for all $\Vr \in \Omega^{I'}$, so that the resulting
$(K_i \ast f_2)(\Vr)$ are accurate for all $\Vr \in \Omega^I$.

Thus we have all the pieces necessary to compute $T_{\tx{K}}^{\tx{int}}$.
In summary, we do this as follows:

\begin{itemize}
\item Using Eq. (\ref{eqn:fastkf}), we can evaluate $(K_i \ast f_2)(\Vr)$ and
$(K_i \ast g_2)(\Vr)$ for points
$\Vr$ on the boundary of a region $\Omega^{I'}$ that is slightly larger than $\Omega^I$.
\item Using those boundary conditions, the Helmholtz equations (\ref{eqn:helm})
 are solved, yielding $(K_i \ast f_2)(\Vr)$ and $(K_i \ast g_2)(\Vr)$ for
all points $\Vr \in \Omega^{I'}$.
\item Then $(K \ast f_2)(\Vr)$ and $(K \ast g_2)(\Vr)$ are constructed with Eq.
(\ref{eqn:sepconvsum}), and we can evaluate $T_{\tx{K}}^{\tx{int}}$ via Eq.
(\ref{eqn:TKV1}).
\end{itemize}

Ther kernel interaction energy $T_{\tx{K}}^{\tx{int}}$ also gives a small
contribution to the forces on region $II$ atoms near the 1-2 boundary.  By
differentiating Eq. (\ref{eqn:TKint}), we find:
\be
\frac{\partial T_{\tx{K}}^{\tx{int}}}{\partial \VR^{II}_j} &=&
-\int \nabla \rho(\Vr - \VR^{II}_j) \left[ f'_{12}(\Vr)(K \ast G)(\Vr)
\right. \nonumber \\
&\mbox{}& + g'_{12}(\Vr)(K \ast F)(\Vr) + F'(\Vr)(K \ast g_2)(\Vr) \nonumber \\
&\mbox{}& \left. + G'(\Vr) (K \ast f_2)(\Vr) \right] \tx{d} \Vr,\\
(K \ast F)(\Vr) &\equiv& \int K(\Vr-\Vr')F(\Vr') \tx{d}\Vr', \nonumber \\
f'_{12}(\Vr) &\equiv& f'(\rho^I(\Vr)+\rho^{II}(\Vr)), \nonumber \\
F'(\Vr) &\equiv& f'_{12}(\Vr) - f'(\rho^{II}(\Vr)), \mbox{ etc.} \nonumber
\ee

\begin{acknowledgements}
The authors wish to thank Paul Maragakis, Ryan Barnett, Emily Carter, 
and William Curtin for fruitful discussions.
This work was supported in part by a MURI-AFOSR Grant No. F49620-99-1-0272.
\end{acknowledgements}


\begin{thebibliography}{10}

\bibitem{WeinanH02}
W. E and Z.~Y. Huang, J. Comput. Phys. {\bf 182},  234  (2002).

\bibitem{CaidBY00}
W. Cai, M. de~Koning, V.~V. Bulatov, and S. Yip, Phys. Rev. Lett. {\bf 85},
  3213  (2000).

\bibitem{HohenbergK64}
P. Hohenberg and W. Kohn, Phys. Rev. B {\bf 136},  B864  (1964).

\bibitem{KohnS65}
W. Kohn and L.~J. Sham, Phys Rev {\bf 140},  1133  (1965).

\bibitem{Gaoreview96}
J. Gao, {\em Reviews in Computational Chemistry}, edited by K.~B. Libkowitz and
  D.~B. Boyd (VCH, New York, N. Y., 1996), pp.\ 119--185.

\bibitem{GovindWdC98}
N. Govind, Y.~A. Wang, A.~J.~R. da~Silva, and E.~A. Carter, Chem. Phys. Lett.
  {\bf 295},  129  (1998).

\bibitem{BroughtonABK99}
J.~Q. Broughton, F.~F. Abraham, N. Bernstein, and E. Kaxiras, Phys. Rev. B {\bf
  60},  2391  (1999).

\bibitem{CuiEKFK01}
Q. Cui {\it et~al.}, J. Phys. Chem. B {\bf 105},  569  (2001).

\bibitem{Daw89}
M.~S. Daw, Phys. Rev. B {\bf 39},  7441  (1989).

\bibitem{DawB84}
M.~S. Daw and M.~I. Baskes, Phys. Rev. B {\bf 29},  6443  (1984).

\bibitem{ErcolessiPT88}
F. Ercolessi, M. Parrinello, and E. Tosatti, Philos. Mag. A-Phys. Condens.
  Matter Struct. Defect Mech. Prop. {\bf 58},  213  (1988).

\bibitem{JacobsenNP87}
K.~W. Jacobsen, J.~K. Norskov, and M.~J. Puska, Phys. Rev. B {\bf 35},  7423
  (1987).

\bibitem{WesolowskiW93}
T.~A. Wesolowski and A. Warshel, J. Phys. Chem. {\bf 97},  8050  (1993).

\bibitem{Cortona91}
P. Cortona, Phys. Rev. B {\bf 44},  8454  (1991).

\bibitem{ChaconAT85}
E. Chacon, J.~E. Alvarellos, and P. Tarazona, Phys. Rev. B {\bf 32},  7868
  (1985).

\bibitem{GarciaGonzalezAC96}
P. Garcia-Gonzalez, J.~E. Alvarellos, and E. Chacon, Phys. Rev. B {\bf 53},
  9509  (1996).

\bibitem{WangT92}
L.~W. Wang and M.~P. Teter, Phys. Rev. B {\bf 45},  13196  (1992).

\bibitem{WangGC98}
Y.~A. Wang, N. Govind, and E.~A. Carter, Phys. Rev. B {\bf 58},  13465  (1998).

\bibitem{WangGC99}
Y.~A. Wang, N. Govind, and E.~A. Carter, Phys. Rev. B {\bf 60},  16350  (1999).

\bibitem{KlunerGWC02}
T. Kluner, N. Govind, Y.~A. Wang, and E.~A. Carter, Phys. Rev. Lett. {\bf 88},
  art. no.  (2002).

\bibitem{ErcolessiA94}
F. Ercolessi and J.~B. Adams, Europhys. Lett. {\bf 26},  583  (1994).

\bibitem{CholyK02}
N. Choly and E. Kaxiras, Solid State Commun. {\bf 121},  281  (2002).

\bibitem{WangBook00}
Y.~A. Wang and E.~A. Carter,  in {\em Theoretical Methods in Condensed Phase
  Chemistry}, {\em Progress in Theoretical Chemistry and Physics}, edited by
  S.~D. Schwartz (Kluwer, Dordrecht, The Netherlands, 2000), pp.\ 117--184.

\bibitem{LiuN89}
D.~C. Liu and J. Nocedal, Math. Program. {\bf 45},  503  (1989).

\bibitem{Hellmann37}
H. Hellmann, {\em Einf\"urung in die Quantenchemie} (Deuticke, Leipzig, 1937).

\bibitem{GoodwinNH90}
L. Goodwin, R.~J. Needs, and V. Heine, J. Phys.-Condes. Matter {\bf 2},  351
  (1990).

\bibitem{FreundN91}
R.~W. Freund and N.~M. Nachtigal, Numer. Math. {\bf 60},  315  (1991).

\end{thebibliography}
\end{document}